\documentclass[twocolumn]{aastex62}

\turnoffedit

\shorttitle{M83 CO PDF}
\shortauthors{Egusa et al.}

\begin{document}

\title{Molecular Gas Properties in M83 from CO PDFs}

\correspondingauthor{Fumi Egusa}
\email{fumi.egusa@nao.ac.jp}


\author{Fumi Egusa}
\affiliation{Institute of Space and Astronautical Science, Japan Aerospace Exploration Agency, Sagamihara, Kanagawa, Japan}
\affiliation{Chile Observatory, National Astronomical Observatory of Japan, Mitaka, Tokyo, Japan}

\author{Akihiko Hirota}
\affiliation{Chile Observatory, National Astronomical Observatory of Japan, Mitaka, Tokyo, Japan}
\affiliation{Joint ALMA Observatory, Santiago, Region Metropolitana, Chile}

\author{Junichi Baba}
\affiliation{Research Center for Space and Cosmic Evolution, Ehime University, Matsuyama, Ehime, Japan}
\affiliation{National Astronomical Observatory of Japan, Mitaka, Tokyo, Japan}

\author{Kazuyuki Muraoka}
\affiliation{Department of Physical Science, Osaka Prefecture University, Sakai, Osaka, Japan}



\begin{abstract}
 We have obtained $^{12}$CO(1--0) data of the nearby barred spiral galaxy M83 
from Atacama Large Millimeter/submillimeter Array and 
Nobeyama 45m observations.
 By combining these two data sets, the total CO flux has been recovered, 
and a high angular resolution ($2''$ corresponding to $\sim 40$ pc at the distance of M83) 
has been achieved.
 The field of view is $3'$ corresponding to $\sim 3.4$ kpc 
and covers the galactic center, bar, and spiral arm regions.
 In order to investigate how these galactic structures affect gas properties, 
we have created a probability distribution function (PDF) of the CO
integrated intensity ($I_{\rm CO}$), peak temperature, and velocity dispersion 
for a region with each structure. 
 We find that the $I_{\rm CO}$ PDF for the bar shows 
a bright-end tail while that for the arm does not.
 Since the star formation efficiency is lower in the bar, 
this difference in PDF shape is contrary to the trend in Milky Way studies where 
the bright-end tail is found for star-forming molecular clouds. 
 While the peak temperature PDFs are similar for bar and arm regions, 
velocity dispersion in bar is systematically larger than in arm.
 This large velocity dispersion is likely a major cause of the bright-end tail and 
of suppressed star formation.
 We also investigate an effect of stellar feedback to PDF profiles 
and find that the different $I_{\rm CO}$ PDFs between bar and arm regions 
cannot be explained by the feedback effect, at least at the current spatial 
\edit1{scale.}
\end{abstract}




\section{Introduction} \label{sec:intro}

 The evolution of interstellar medium (ISM) in galactic disks is intimately related to 
star formation and galaxy evolution.
 In external face-on galaxies, compared to Milky Way (MW), 
it is easier to investigate ISM conditions 
and their evolution while rotating in the disk.
 Recent observations have reported that ISM conditions depend on 
associated galactic structures, such as spiral arms and bars.
 For example, massive clouds tend to reside in spiral arms 
\citep{Koda09,Colo14a}, suggesting that 
spiral arms play an important role to form 
massive clouds that are plausible precursors of massive star clusters.
 Line ratios have also been found to be different among disk structures 
\citep[e.g.][]{Koda12b,Tosa17}, representing different physical conditions.
 High resolution observations revealed ISM evolution within spiral arms 
\citep{Egu11,Hiro11}, clouds at the downstream side of spiral arms 
being more massive and star-forming.
 In bar regions, molecular gas is concentrated in ridges and bar edges \citep[e.g.][]{Sheth02}
but star formation efficiency is relatively low especially in ridges 
\citep[e.g.][]{Momo10,Sor12,Pan17}.
 Recent numerical calculations have also suggested that 
ISM conditions depend on dynamical environments 
\edit1{\citep[e.g.][]{FujiY14,Ngu17}.}


 Many of studies mentioned above defined clouds 
using cloud identifying algorithms.
 One caveat of such algorithms is that they tend to identify structures 
whose size is similar to an angular resolution.
 As a result, a dynamic range of cloud size is rather small.
 In addition, recent numerical simulations suggest that 
molecular clouds are not isolated and discrete structures 
but are part of continuous fluid \citep[e.g.][]{Dob13}.
 For example, \citet{Colo14a} identified $>1500$ clouds in the nearby spiral galaxy M51 
from a CO(1--0) data set at 40 pc resolution 
and most of the clouds have a radius between 20 pc and 100 pc 
(i.e.\ the dynamic range is $\sim 1$ dex, their Figure 8).
 They also calculated the mass in the clouds to be about half of 
the total mass.
 The cloud analysis cannot take into account such gas outside identified clouds 
whose mass can be comparable to that inside the clouds.

 Another and complementary way of measuring ISM properties is to use a 
probability distribution function (PDF), which is a normalized histogram of a physical parameter
or an observed quantity.
 Advantages of PDF analysis are that it is free of any biases due to cloud identification 
and that it includes ISM components which would not be included in identified clouds.

 For Milky Way (MW) molecular clouds, the PDF of gas column density has been used to investigate 
their star-forming conditions.
 \citet{Kai09} created gas column density PDFs based on NIR extinction maps
and found that those for star-forming clouds show a power-law (PL) like tail 
while those for quiescent clouds show a log-normal (LN) profile.
 Such different PDF profiles have been interpreted as different ISM conditions 
-- LN for turbulence-dominated clouds \citep[e.g.][]{NP99} and 
PL for self-gravity-dominated clouds \citep[e.g.][]{Kri11,Giri14} 
or clouds with high density contrast \citep{Elm11a}.
 \citet{Brunt15} claimed that the tail can be explained 
as a part of a wide LN profile which corresponds to a cold gas component.

 Brightness distribution function (BDF) introduced by \citet{Sawa12a} 
is a PDF of CO brightness temperature.
 Together with their following study \citep{Sawa12b}, 
they found that BDFs for spiral arms in MW tend to show a bright-end tail
compared to interarm regions.
 Meanwhile, \citet{Rath14} presented that a cloud close to the MW center 
shows a PDF peaking at a higher density 
compared to clouds in the solar neighborhood.
 The PDF peak corresponds to a mean density and 
the authors attributed this difference to high-pressure environments around the MW center.

 For external galaxies, \citet{Hugh13a} created PDFs for galactic structures of M51 
from CO(1--0) data at $\sim 40$ pc scale and found different PDF shapes among the structures.
 The difference is more clearly seen in PDFs of the integrated CO intensity 
in the sense that center and arm PDFs are wider and deviate from LN profiles 
while interarm PDFs are narrower and close to LN profiles.
 Interestingly, PDFs for the center and arm where star formation is relatively active 
appear to be truncated rather than extended.
 They interpreted this truncation as a feedback effect.
 They also claimed that the narrow LN profiles in the interarm 
are a sign of turbulence dominated ISM.
 \citet{Dru14} created a PDF at $\sim 50$ pc scale for M33 based on CO(2--1) integrated intensity
and found that its shape positively deviates from the LN profile 
at $N({\rm H_2}) > 2 \times 10^{21}~{\rm H_2/cm^2}$.
 They attributed this bright-end excess to self-gravity, 
although the threshold column density is low 
(comparable to the nominal 3$\sigma$ limit of M51 data by \citet{Hugh13a}).
 We should note here that spatial resolutions of these studies are 
much larger than those of MW studies.
 Thus, a direct comparison of MW and these studies is not straightforward.

 \citet{Hugh13a} also investigated how data quality (i.e.\ resolution and sensitivity) and 
technique to create integrated intensity maps affect PDF shapes.
 Their results suggest that the high spatial resolution is important to 
investigate bright-end profiles.
 The importance of spatial resolution is also pointed out by 
numerical simulations in \citet{Wada07}.
 Consistently, \citet{Ber15} found that PDFs of gas surface densities 
for M31 and M51 at $\sim 300$ pc scale 
are well fitted by LN profiles.
 \citet{Oss16} investigated how observing conditions affect PDF shapes.
 In particular, they presented that interferometric observations should distort 
overall PDF shapes, i.e.\ their faint-end profile, peak position, and bright-end profile.
 To minimize the distortion, a dense uv sampling especially at low spatial frequencies 
is necessary.
 Given this dependency, we compare PDF profiles only within a galaxy 
and do not directly compare the results with other galaxies.


 In this paper, we present CO PDFs for  
the nearby \citep[$D=4.5$ Mpc;][]{Thim03} 
face-on \citep[inclination angle = $24^\circ$;][]{Com81} 
barred spiral \citep[SAB(s)c;][]{RC3} galaxy M83 (or NGC 5236).
 Its proximity and disk orientation are well suited to study 
ISM properties and their relationship to disk structures. 
 We find that PDFs differ significantly among the galactic structures
and discuss the origin of this difference.
 Stellar feedback effects on PDF profiles are also investigated.
 This is a complementary study to Hirota et al.\ 
 (\edit1{submitted}, hereafter paper I), 
which investigates cloud properties based on the same data set.


\section{Observations and data reduction}
 The rotational transitions of CO are the most widely used tracers of molecular gas, 
and their integrated intensities have been used to estimate the column density of molecular hydrogen
\citep[e.g.][]{YS91,Bola13b}.
 We have mapped a central disk area of M83 
in the $^{12}$CO(1--0) emission using Atacama Large Millimeter/submillimeter Array 
(ALMA) and the Nobeyama 45m telescope (NRO45m).
 A detail of observations and data reduction will be in paper I, and 
we provide a summary here.

\subsection{ALMA observations}
 The data have been obtained for the Cycle 0 project 2011.0.00772.S, 
and consist of two Execution Blocks (EBs) with a compact configuration and 
three EBs with an extended configuration.
 The field of view (FoV) is about $2.6' \times 2.6'$ with 
45 mosaic pointings, which includes the galactic center, 
a northern part of the bar, and a spiral arm.

\subsection{Nobeyama 45m observations}
 The NRO45m observations were carried out 
using the on-the-fly mode with the T100 receiver and SAM45 spectrometers.
 The entire FoV includes the ALMA FoV.

\edit1{
\subsection{Data combination}\label{sec:combine}
}
 In order to combine the ALMA and NRO45m data, 
the NRO45m map is first converted to a uv data set.
 Then, the two uv data sets from ALMA and NRO45m 
are Fourier-transformed and deconvolved together.
 
 The synthesized beam is $2.0''\times 1.1''$ in full width at half maximum.
 At the distance of M83, this corresponds to 45 pc $\times$ 24 pc.
 We adopt a $1\sigma$ sensitivity of each channel to be 8.0 mJy/beam 
or 0.33 K with a channel width of 2.6 km/s.
 The pixel size is set to be $0.25''$.

 The moment 0 (i.e.\ integrated intensity; $I_{\rm CO}$) 
and moment 2 (i.e.\ velocity dispersion; $\sigma_V$) maps are
created by 
\edit1{
applying the mask that includes
all the pixels with S/N $\ge 4$ and 
those with S/N $\ge 2$ and morphologically connected to the former pixels.
 The noise map is not uniform according to the primary beam pattern 
and to different observing conditions (e.g.\ integration time) between mosaic pointings.
}
 A typical $1\sigma$ sensitivity of the integrated intensity map is estimated to be 
47 mJy/beam km/s or 1.8 K km/s.
 Assuming the Galactic conversion factor, 
$X_{\rm CO} = 2\times 10^{20}~{\rm H_2/cm^2/(K~km/s)}$ \citep{Dame01}, 
this sensitivity corresponds to $\Sigma = 3.6\times 10^{20}~{\rm H_2/cm^2}$
or $5.8~M_\odot/{\rm pc^2}$.
 Note that this is not corrected for inclination.

\section{CO PDFs}\label{sec:pdf}
 In this section, we present PDFs of the integrated intensity ($I_{\rm CO}$), 
the peak temperature ($T_{\rm peak}$), and the velocity dispersion ($\sigma_V$), 
and discuss their profiles according to disk structures (\S \ref{sec:disk}) 
and to stellar feedback (\S \ref{sec:feedback}).
\edit1{
 Note that $\sigma_V$ is not corrected for the galactic rotation and beam smearing.
 As described in Appendix,
these effects do not affect the following results and discussions.
}



\subsection{Disk structures}\label{sec:disk}
 According to the Fourier analysis of $K_{\rm s}$-band image by 
\citet[hereafter H14]{Hiro14}, 
we define the radial range of the bar to be $r=20''$--$85''$.
 This is the region where non-circular motions are clearly seen in 
a position-velocity diagram (their Figure 7).
 The semi-minor axis of the bar is set to be $46''$ 
based on an ellipse fitting to the same $K_{\rm s}$-band image.
 A region outside the bar, mostly consists of a spiral arm, is called an arm region.
 A region inside the bar is called a center region.
 The center and bar regions are indicated by dots-dashed and dashed lines 
in Figure \ref{fig:Ico_disk} (left), respectively.
 The definition in this paper is different and more simple compared to that of paper I, 
for the sake of better statistics especially for PDF profiles at the bright end.

\subsubsection{$I_{\rm CO}$ PDF}

\begin{figure*}[h]
\includegraphics[width=0.4\linewidth]{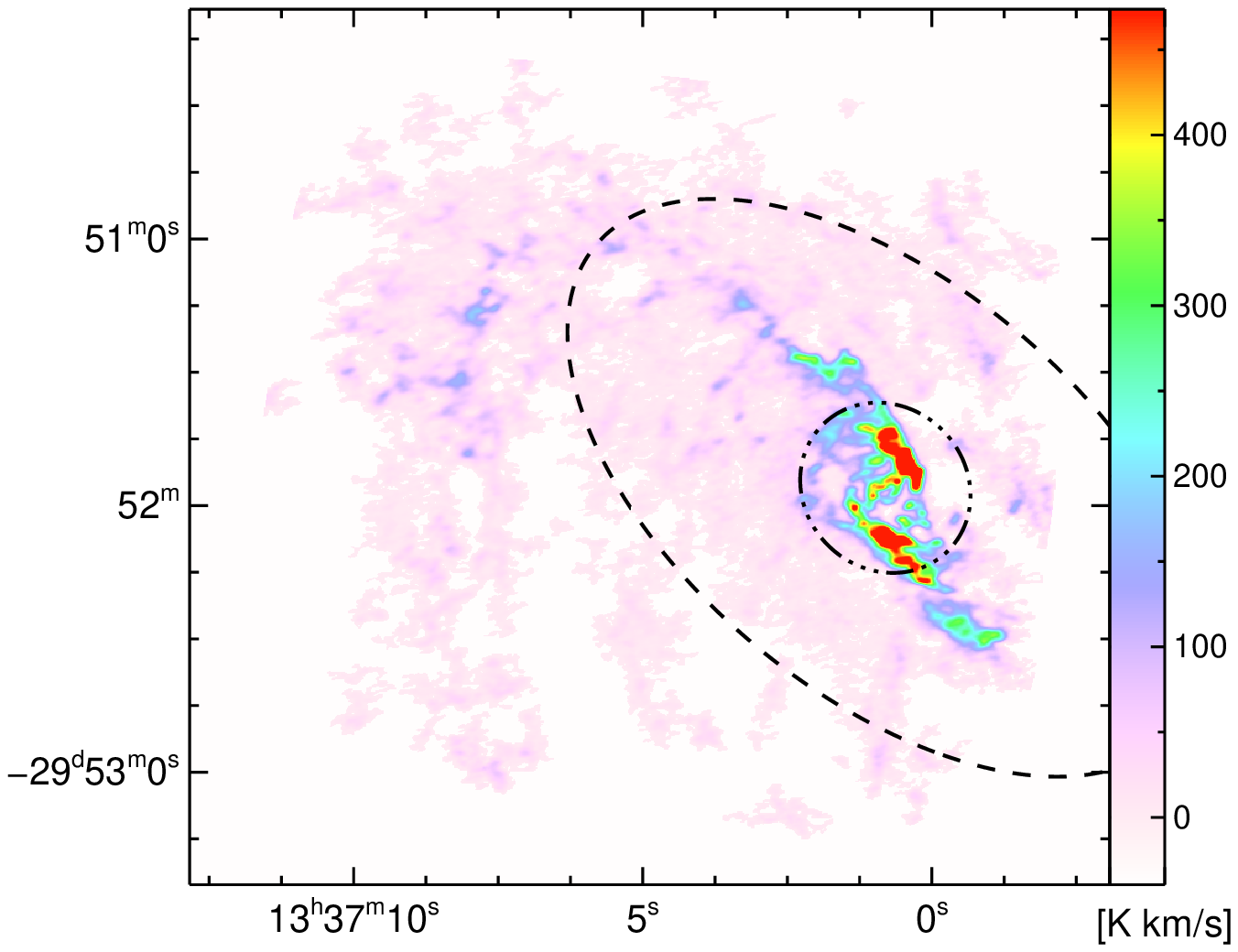}
\includegraphics[width=0.4\linewidth]{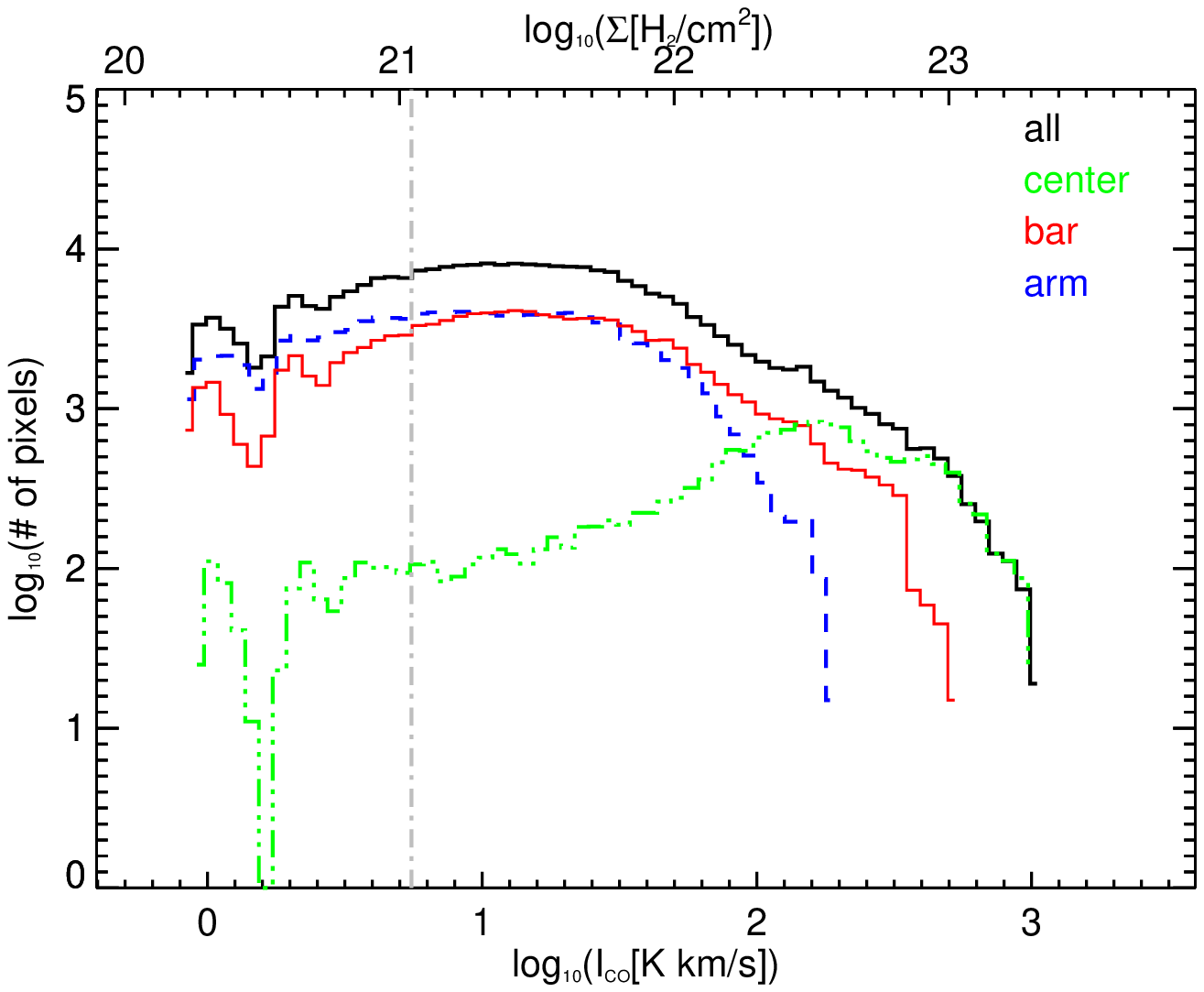}
\includegraphics[width=0.19\linewidth]{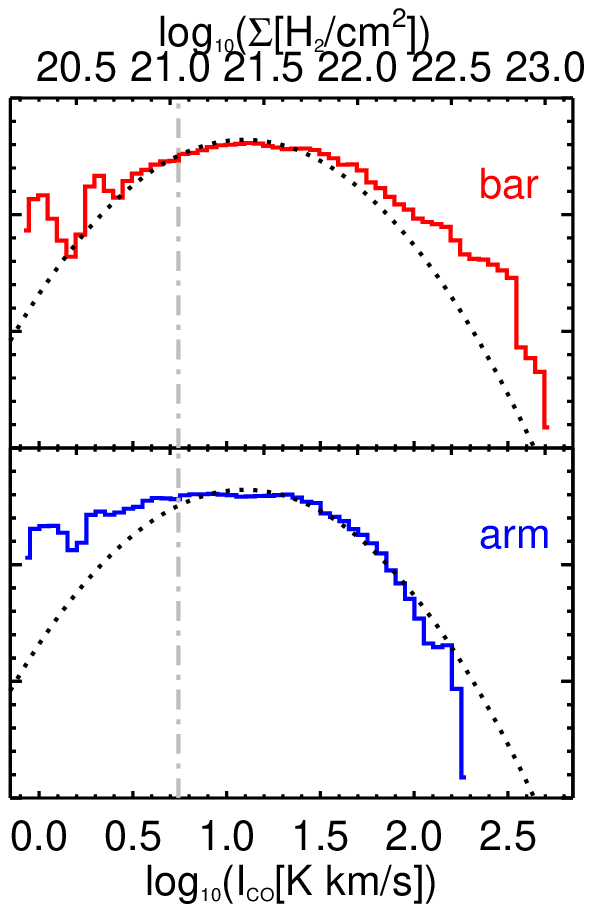}
\caption{{\it Left:} The $I_{\rm CO}$ map of M83 from ALMA+NRO45m observations.
The coordinates are Right Ascension and Declination in ICRS.
The outer boundary of center and bar regions are indicated by 
3dots-dashed and dashed lines, respectively.
\edit1{{\it Middle:}}
The $I_{\rm CO}$ PDFs for the center (green 3dots-dashed), 
bar (red solid), and arm (blue dashed) regions. 
The black solid line is the PDF for the entire FoV.
The vertical gray dot-dashed line indicates the $3\sigma$ of $I_{\rm CO}$. 
The horizontal axis on top is the molecular gas surface density in ${\rm H_2/cm^2}$, 
calculated assuming the Galactic conversion factor.
\edit1{{\it Right:} The same $I_{\rm CO}$ PDFs for the bar and arm 
with an LN function (black dotted line).}
}
\label{fig:Ico_disk}
\end{figure*}

 The $I_{\rm CO}$ map and its PDF for each of the three regions 
are presented in Figure \ref{fig:Ico_disk}.
 Peak positions of bar and arm PDFs are similar (10--30 K km/s), while that for center 
is an order of magnitude higher.
 This difference in peak position is similar to what is found for MW clouds \citep{Rath14}.
 We have confirmed that this general trend holds when the outer boundary of 
the center region is changed to $r=10''$ or $30''$.

 The arm PDF is likely LN or at least without a clear PL tail.
 Its shape as well as the dynamic range of gas surface density are 
similar to those of spiral arms in M51 derived by \citet{Hugh13a}.
\edit1{
 In the rightmost panel of Figure \ref{fig:Ico_disk}, 
an LN function fitted to the arm PDF above $3\sigma$ (the gray dot-dashed vertical line)
is presented by the black dotted line.
 The arm PDF appears to be slightly truncated at the bright end, 
but the fit result highly depends on the range used for the fit.
}
 Furthermore, as already mentioned in \S \ref{sec:intro}, 
\citet{Oss16} presented that PDF profiles are highly sensitive to 
the data quality -- not only the sensitivity but also the uv coverage.
 In this study, we thus 
\edit1{
use this LN profile only for the purpose of guiding the eye, 
and do not compare the shape of profiles
}
with other studies.

 On the other hand, PDFs for the center and bar regions 
extend to higher densities 
\edit1{compared to the arm PDF.}
 The center PDF appears to have two peaks, 
while the bar PDF exhibits a PL like tail at the bright end ($I_{\rm CO} \ga 100$ K km/s).
\edit1{
 The excess is also clear when compared to the LN profile 
(Figure \ref{fig:Ico_disk} (right)).
}
 As seen in Figure \ref{fig:Ico_disk} (left), such bright emission comes from 
the molecular ridges close to the center region 
and not so much gas concentration is found in the northern bar end.
 This is consistent with previous CO observations of M83, 
where gas concentration is found only at the southern bar end (H14), 
which is not covered by the current FoV.

 From PDF studies of MW clouds, 
a column density PDF with a bright-end tail has been regarded 
as a sign of star formation \citep{Kai09}. 
 However, star formation efficiency (SFE = star formation rate / (molecular) gas mass) in the bar of M83 is 
lower than that in the arms (H14).
 One reason for this difference would be the difference in spatial scales. 
 While MW studies discuss PDF profiles at pc scale within a cloud, 
M83 PDFs in this paper are at 40 pc scale, which is comparable to a typical MW cloud size.
 Results presented here indicate that at this scale, 
an existence of a bright-end tail of an $I_{\rm CO}$ PDF 
is not always the sign of active star formation.
 In the following subsections, we further discuss possible explanations for this finding.

\subsubsection{$T_{\rm peak}$ and $\sigma_V$ PDF}\label{sec:TV_disk}
\begin{figure*}[h]
\includegraphics[width=0.4\linewidth]{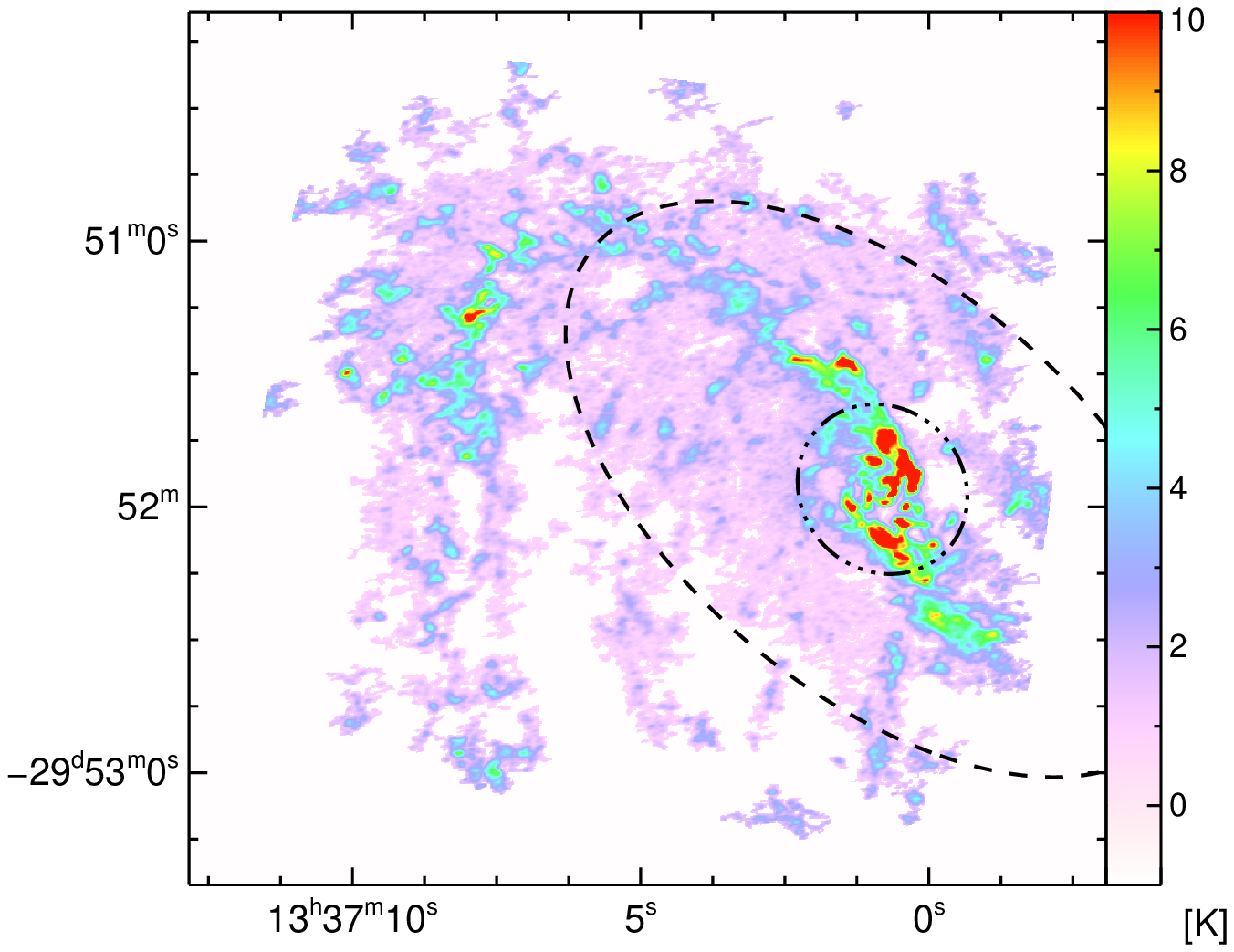}
\includegraphics[width=0.4\linewidth]{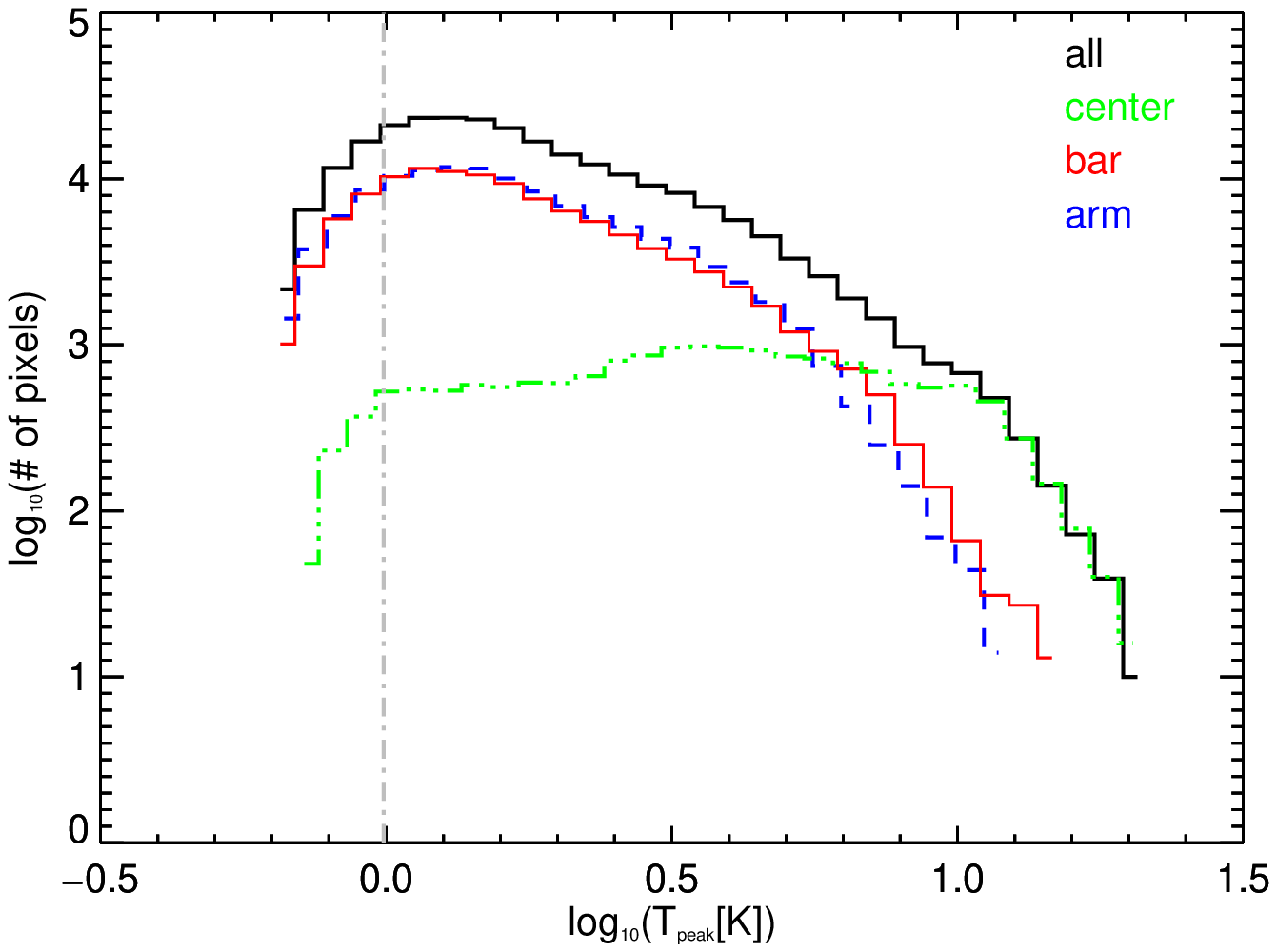}
\caption{
Same as Figure \ref{fig:Ico_disk}, but for $T_{\rm peak}$.
The vertical gray dot-dashed line indicates the $3\sigma$ of channel maps in K.
}
\label{fig:T_disk}
\end{figure*}

\begin{figure*}[h]
\includegraphics[width=0.4\linewidth]{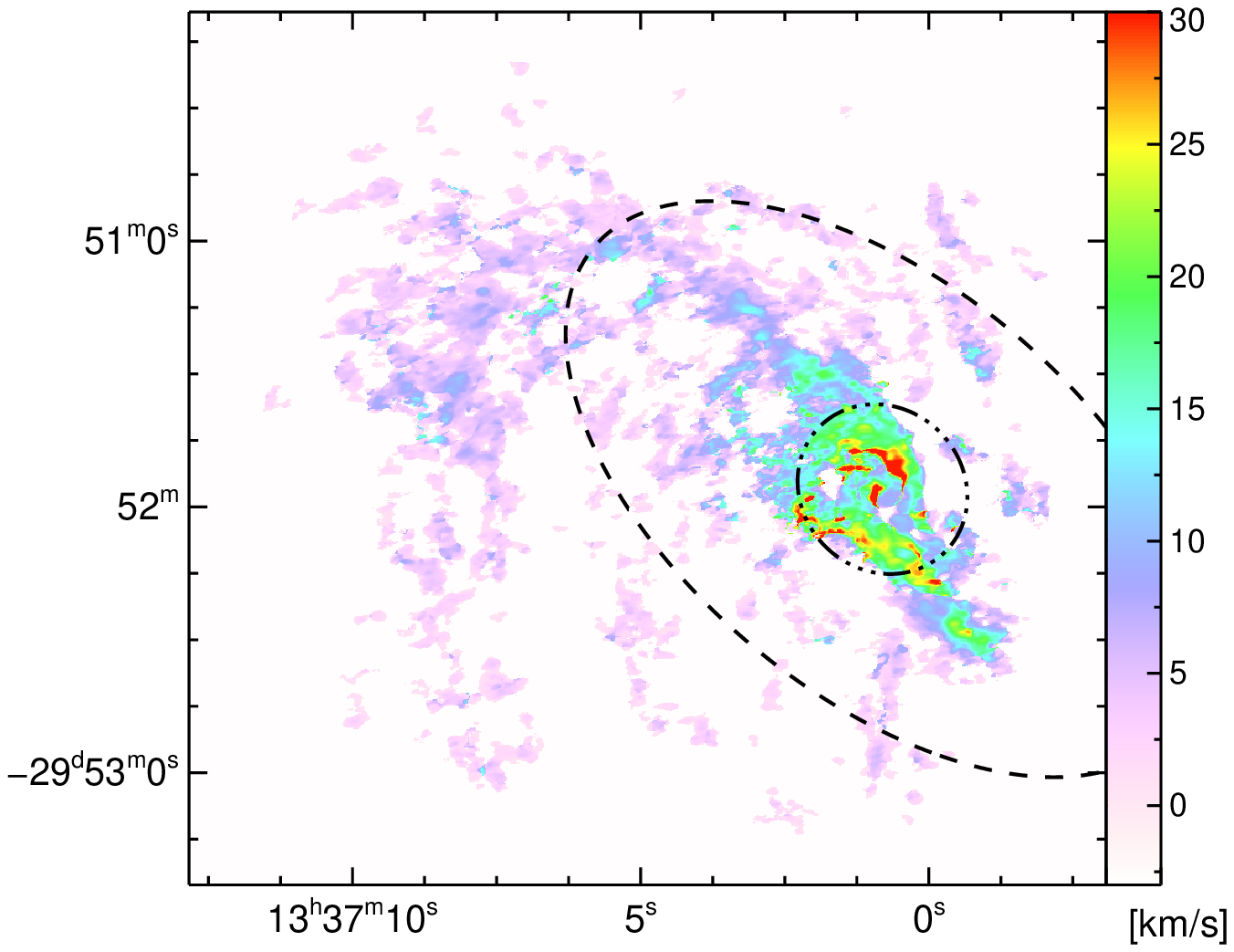}
\includegraphics[width=0.4\linewidth]{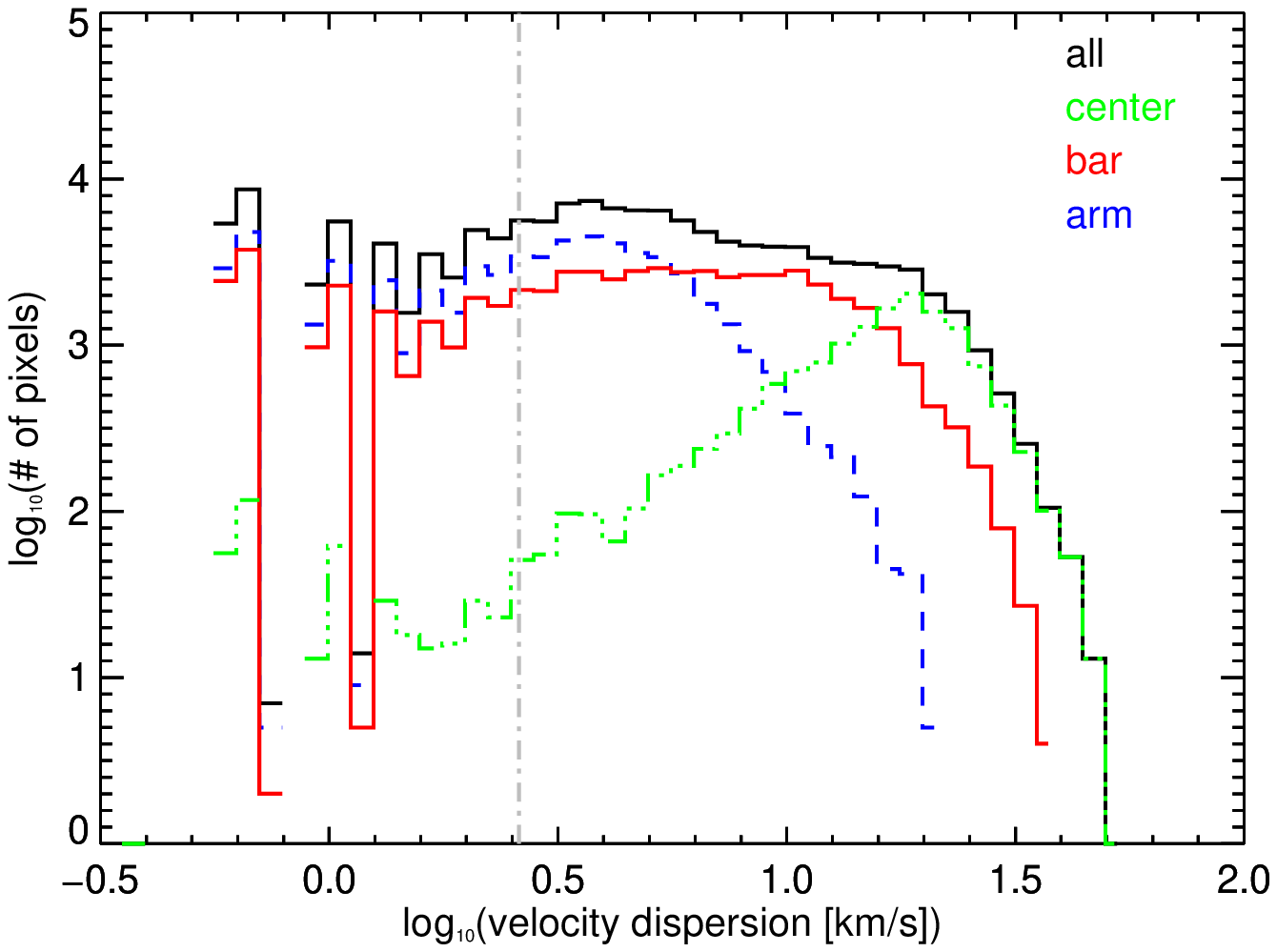}
\caption{
Same as Figure \ref{fig:Ico_disk}, but for $\sigma_V$.
The vertical gray dot-dashed line is the channel width ($2.6$ km/s).
}
\label{fig:dV_disk}
\end{figure*}

 In order to investigate the reason for the difference in $I_{\rm CO}$ PDFs 
between the bar and arm regions, we create PDFs of other observed properties. 
 Figure \ref{fig:T_disk} presents the $T_{\rm peak}$ map and its PDFs, 
while Figure \ref{fig:dV_disk} presents those for $\sigma_V$.
 While $T_{\rm peak}$ PDFs do not differ between the bar and arm regions, 
the $\sigma_V$ PDF for the bar is shifted to the right-hand side of that for the arm.
 In M51, \citet{Hugh13a} found a similar trend that 
their brightness temperature PDFs are more similar between structures
compared to $I_{\rm CO}$ PDFs.
 Since $I_{\rm CO} \sim T_{\rm peak} \times \sigma_V$, 
this large velocity dispersion in the bar region likely results in 
the bright-end tail of the $I_{\rm CO}$ PDF.
 A reason for this large velocity dispersion would be one or combination of 
large shear motions within the bar, 
large random motions within a cloud, 
and overlaps or collisions of clouds at different velocities.
 The second point is supported by large velocity dispersions of molecular clouds in the bar 
presented in paper I.
\edit1{
 A more detailed discussion on how these dynamic properties of molecular gas 
are related to star formation activities is given in \S \ref{sec:discussion}.
}

\edit1{
 For the center region, both $T_{\rm peak}$ and $\sigma_V$ show different 
PDF profiles compared to the other regions.
 The excess in the $I_{\rm CO}$ PDF comes from both high $T_{\rm peak}$ 
and large $\sigma_V$.
 Such high $T_{\rm peak}$ and large $\sigma_V$ are also observed in 
individual clouds identified in paper I.
 The high $T_{\rm peak}$ is a sign of high gas temperature 
due to the circumnuclear starburst \citep[e.g.][]{Knap10,Woff11} and/or a large filling factor.
 The large $\sigma_V$ comes from either large streaming motions 
or highly turbulent clouds.
}


\begin{figure*}[h]
\includegraphics[width=0.4\linewidth]{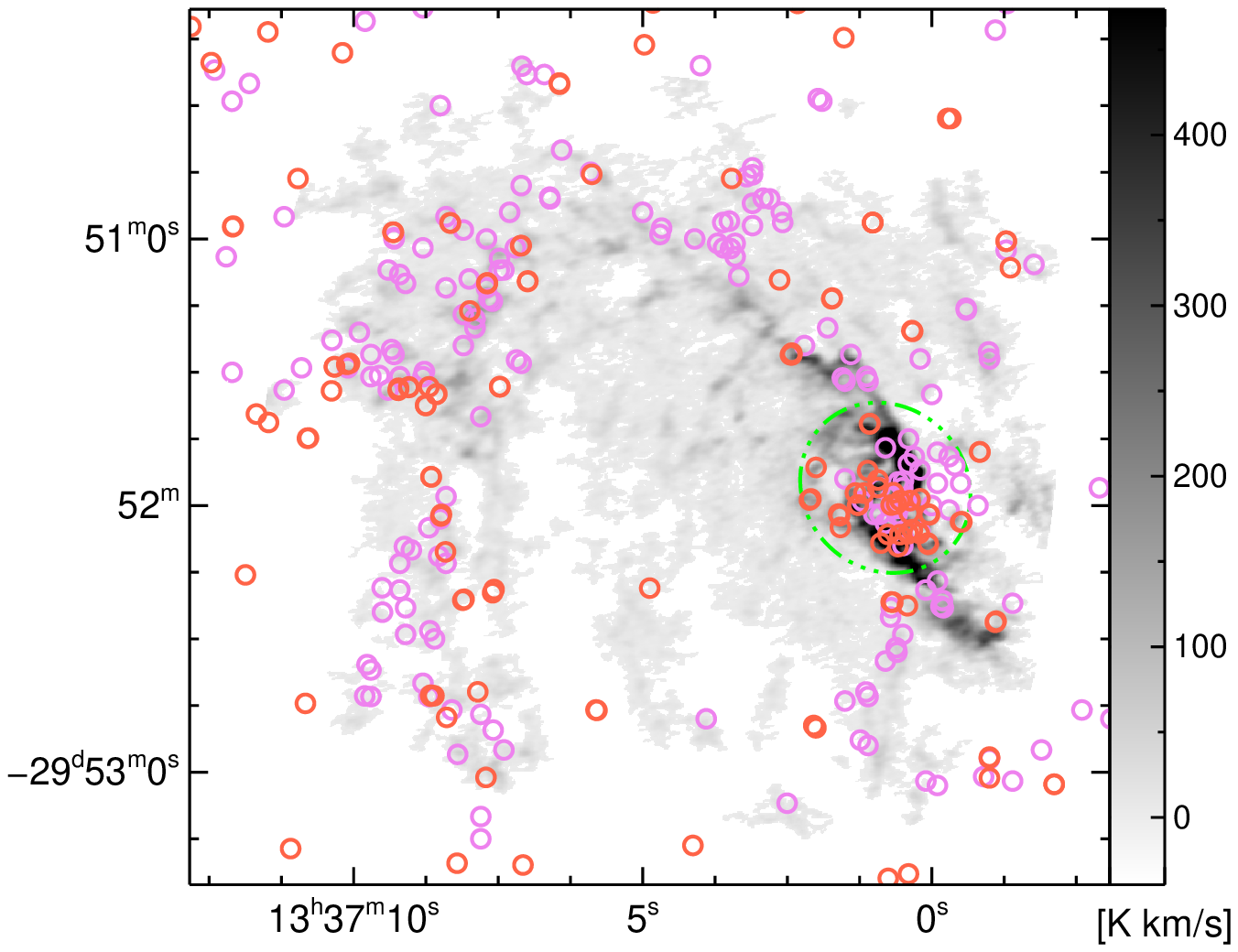}
\includegraphics[width=0.4\linewidth]{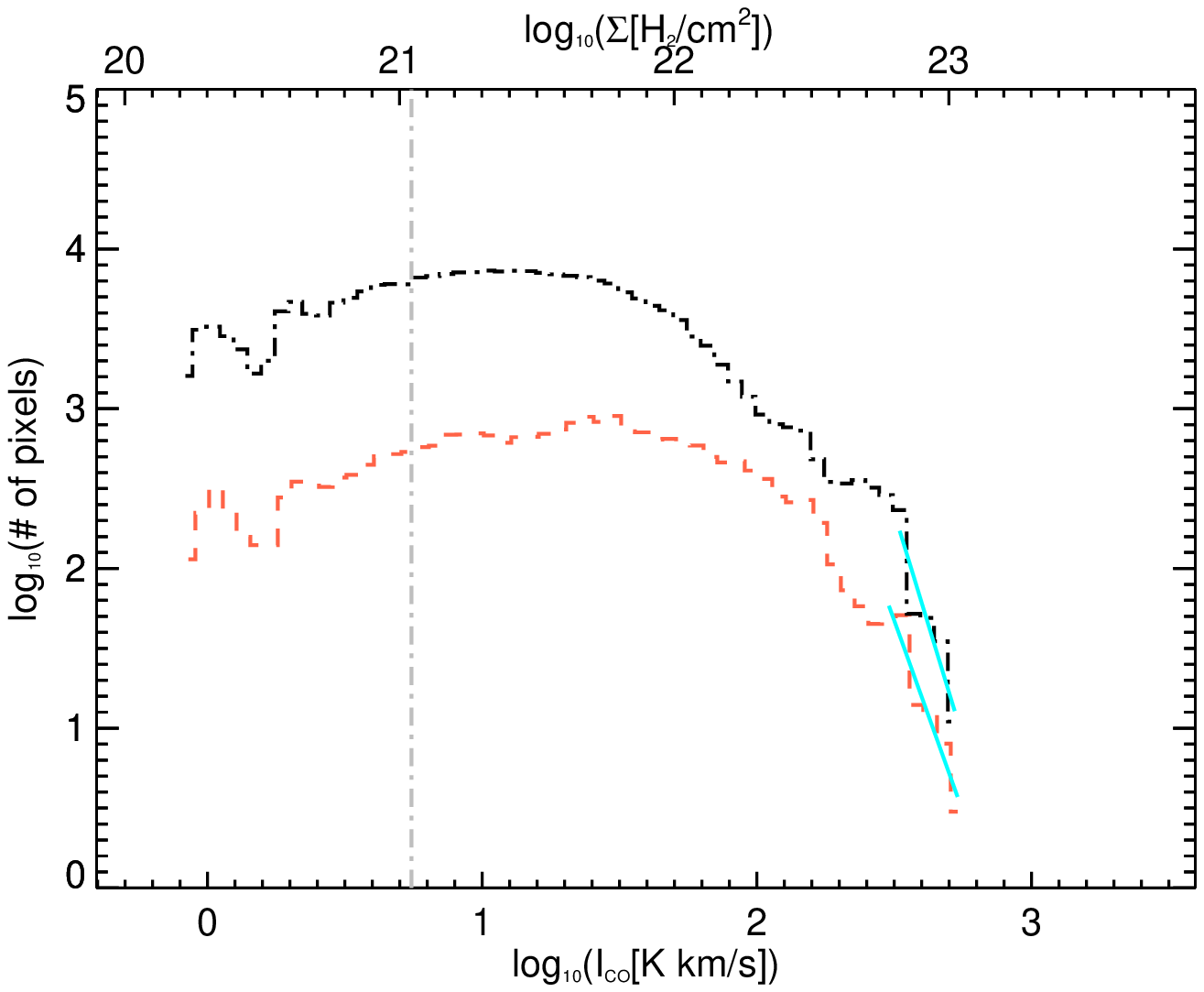}
\caption{
{\it Left:} Circles indicate the positions of 
young stellar clusters or HII regions (purple) 
and SNRs (red) with a radius of $2''$.
The background image is the $I_{\rm CO}$ map (same as Figure \ref{fig:Ico_disk}, but in a different color).
{\it Right:} PDFs for feedback (red dashed) and non-feedback (black \edit1{dot-dashed}) area.
The center region is excluded.
\edit1{Cyan solid} lines indicate the linear fit results at the bright end.
}
\label{fig:Ico_fb}
\end{figure*}
\begin{figure*}[h]
\includegraphics[width=0.4\linewidth]{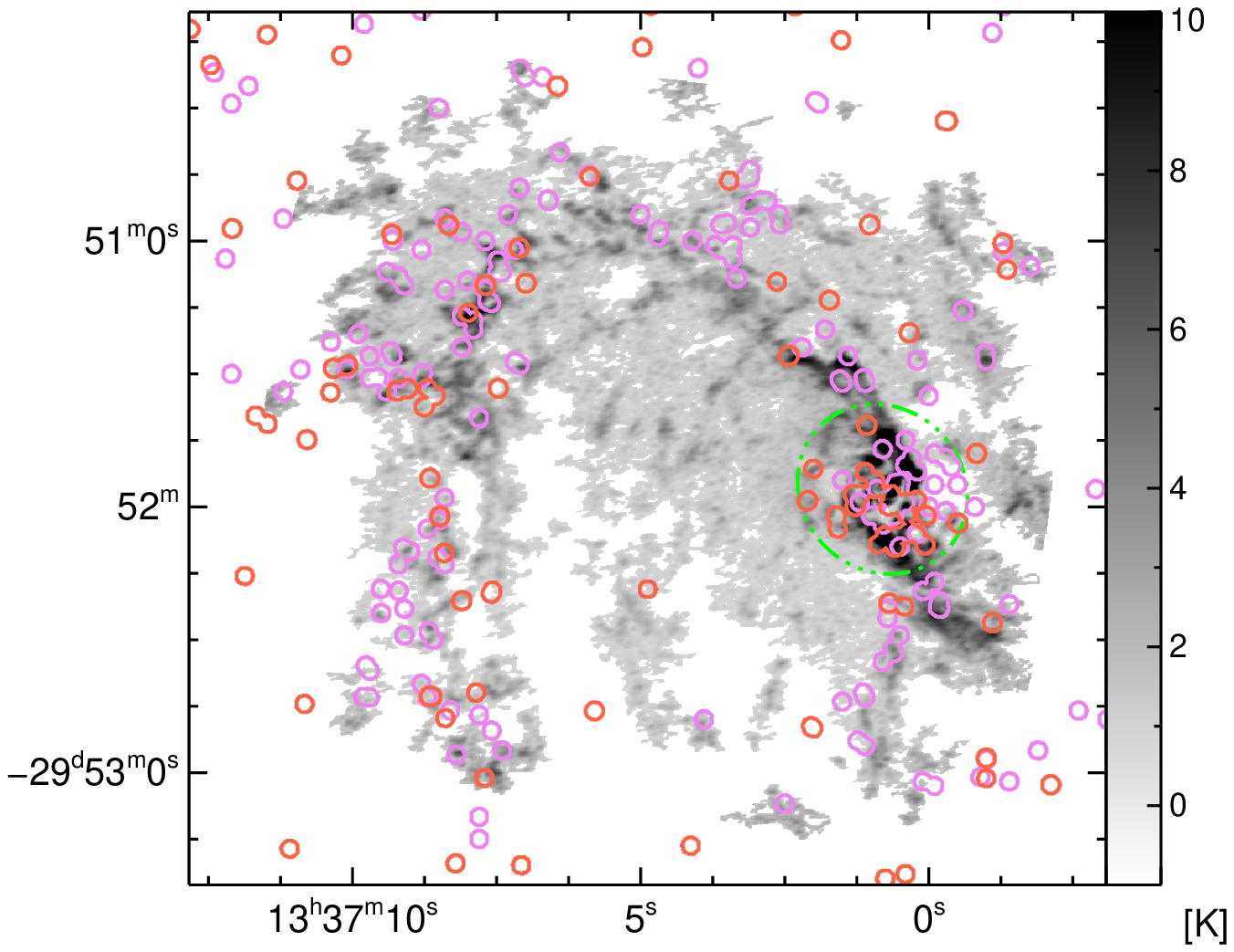}
\includegraphics[width=0.4\linewidth]{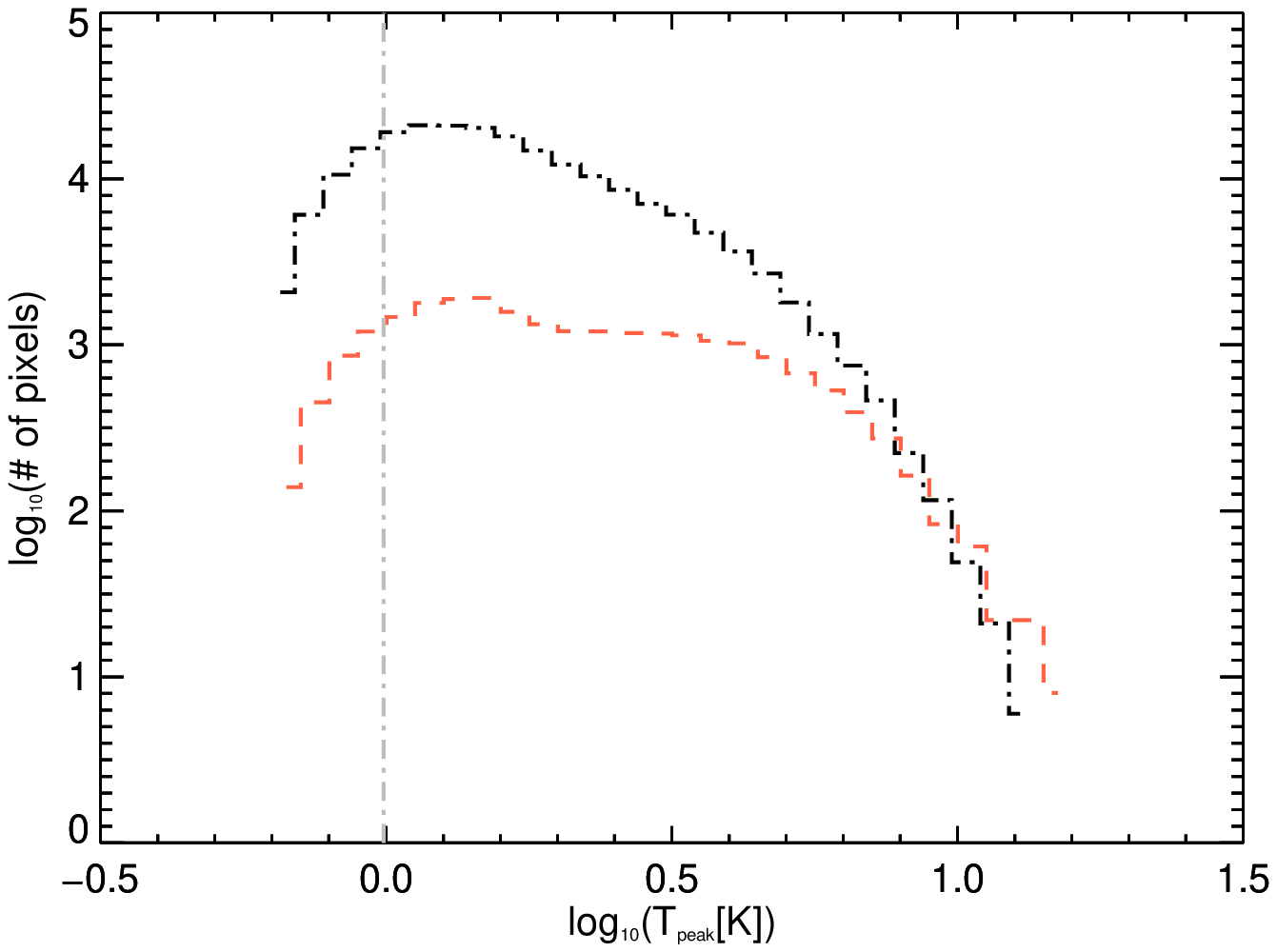}
\caption{
Same as Figure \ref{fig:Ico_fb}, but for $T_{\rm peak}$.
}
\label{fig:T_fb}
\end{figure*}
\begin{figure*}[h]
\includegraphics[width=0.4\linewidth]{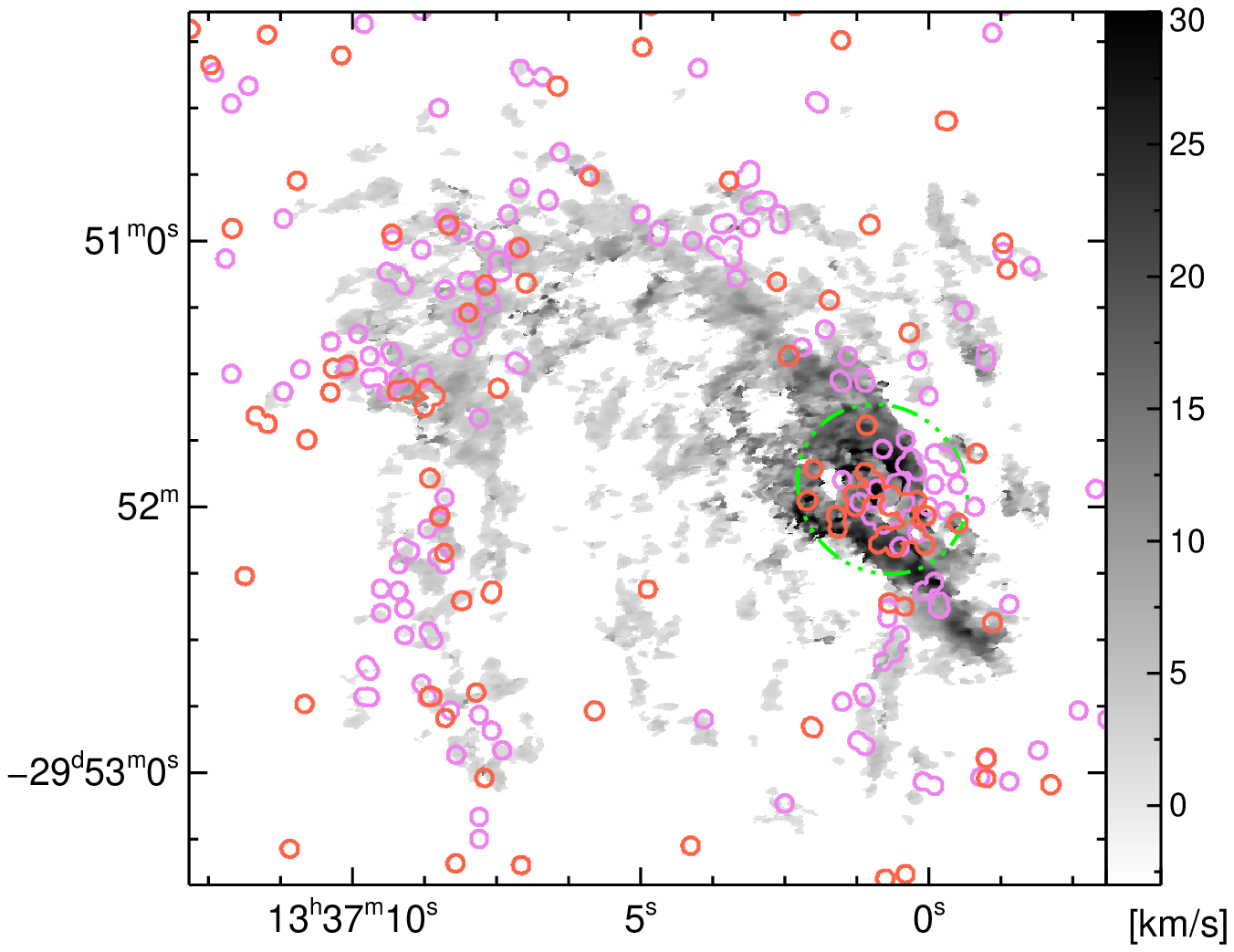}
\includegraphics[width=0.4\linewidth]{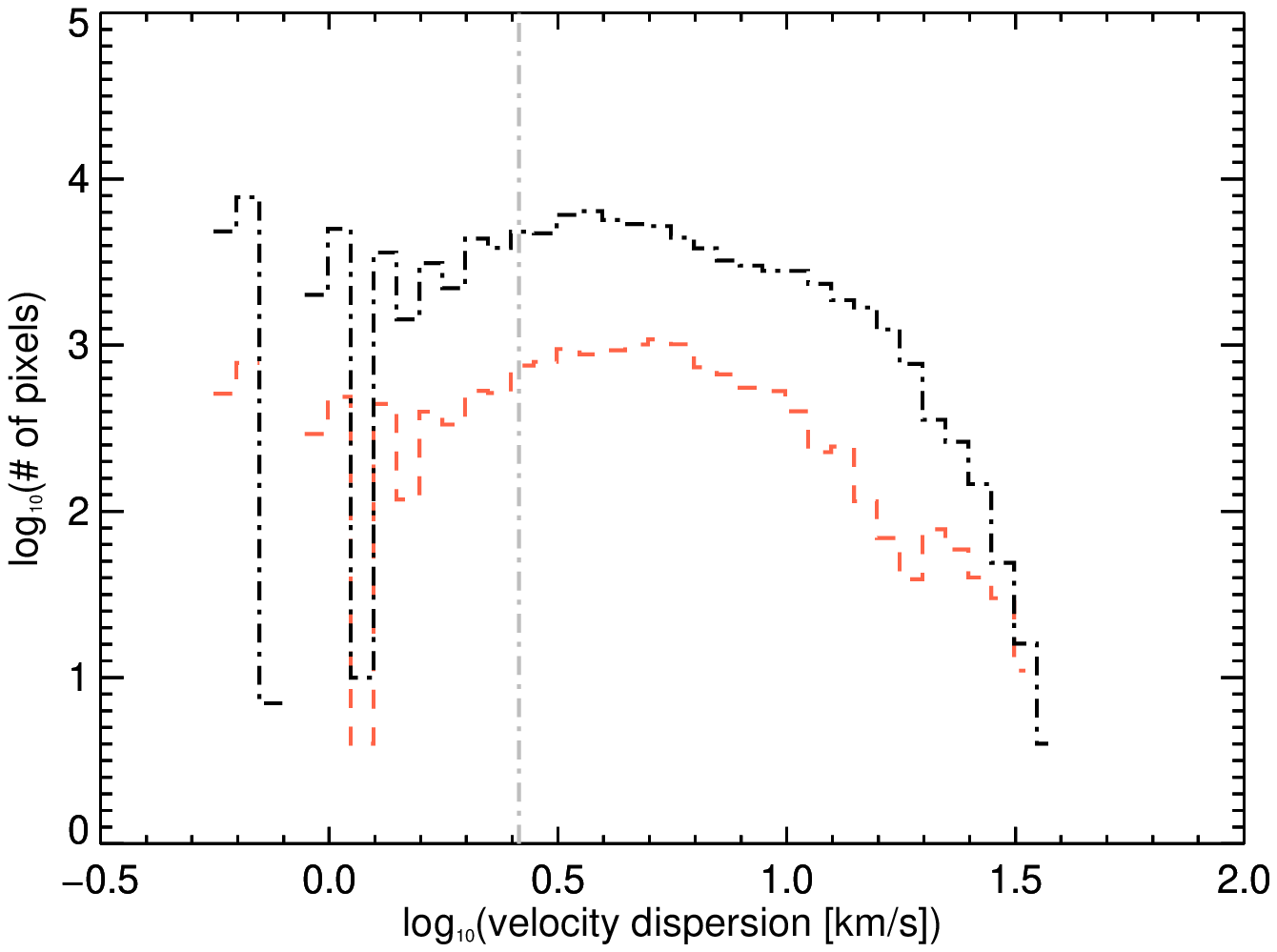}
\caption{
Same as Figure \ref{fig:Ico_fb}, but for $\sigma_V$.
}
\label{fig:dV_fb}
\end{figure*}

\subsection{Stellar feedback}\label{sec:feedback}


 Another explanation for the different $I_{\rm CO}$ PDFs is that 
the arm PDF is truncated due to the effect of stellar feedback 
such as radiation pressure, photoionization, 
and supernova explosion
\citep{Hop12}.
 Such stellar feedback mechanisms can preferentially 
destroy dense gas around young stars, and thus truncate density PDFs 
at the bright end.

 In order to investigate the feedback effect on PDF profiles, 
we have defined ``feedback area'' to be within $2''$ from known 
young star clusters, HII regions, or supernova remnants (SNRs), 
where stronger feedback is expected. 
\edit1{
 The catalogs we use in this paper are 
\citet{Whit11} for young star clusters 
(their categories 3, 4a, and 4b only),
H14 for bright HII regions (sources with $L({\rm H}\alpha) > 10^{37.6}$ erg/s only), 
and \citet{Dop10} and \citet{Blair12,Blair14} for SNR candidates. 
 We have confirmed that $>90$\% of sources within our FoV have radius $<2''$.
 The defined
}
feedback area are indicated by circles in the left panel of Figure \ref{fig:Ico_fb}.

\subsubsection{$I_{\rm CO}$ PDF}
 PDFs for feedback and non-feedback area are presented in 
the right panel of Figure \ref{fig:Ico_fb}.
 Since we are interested in difference between bar and arm regions, 
the center region (enclosed by a dots-dashed line in the left panel of Figure \ref{fig:Ico_fb}) is excluded.
 Contrary to the theoretical expectation, 
the bright-end slopes of 
these two PDFs are not significantly different.
 Based on numerical simulations by \citet{Hop12}, 
we fit each PDF at $\Sigma > 10^3~M_\odot/{\rm pc^2} = 6\times 10^{22}~{\rm H_2/cm^2}$ 
with a line.
\edit1{
The fit is performed in the log-log space using the least chi-square method.
}
 Although the number of data points that exceed this gas density limit is small, 
slopes of the two fitted lines (\edit1{solid cyan} lines in Figure \ref{fig:Ico_fb}) are consistent within errors. 
 We thus deduce that at least at this spatial \edit1{scale (i.e., $\sim 40$ pc in radius)}
the feedback effect is not dominant in shaping the bright-end profiles of $I_{\rm CO}$ PDFs.

\edit1{
 We have also tried different definition of radius ($r=1''$ and $3''$) for the feedback area.
 For the former case, the bright-end slope is steeper for the feedback area 
than that for the non-feedback area.
 However, the feedback PDF has only two data points above the threshold, 
which results in failure of deriving errors.
 For the latter case, the slope is shallower for the feedback area than that for the non-feedback area.
 A naive interpretation of this result is that stellar feedback 
reduces the amount of dense gas within 20 pc and 
increases it at 60 pc away from the source.
 The former is consistent with the general understanding that 
stellar feedback effect should be larger when closer to the source.
 The latter may indicate the positive feedback effect, 
in which the swept-out gas forms the next generation stars \citep[e.g.][]{Deh10}.
 However, this could also be due to the bias that 
the sources are mostly around the spiral arm where 
the gas density is generally high.
 For further discussion, CO data at a higher spatial resolution 
($< 10$ pc considering the size distribution of feedback sources) are essential.
}

\subsubsection{$T_{\rm peak}$ and $\sigma_V$ PDF}
 One thing we should note here that PDFs in the simulations are for 
the gas column density $\Sigma$ while PDFs from observations are for 
the $I_{\rm CO}$.
 Although $X_{\rm CO}$ is often assumed to be constant within a galaxy, 
it could vary especially at smaller scales.
 For example, gas in the feedback area can be heated, 
which results in brighter CO emission.
 In addition, the stellar feedback may accelerate gas, which results in larger velocity dispersion.
 Such positive feedback effects can compensate the negative effect suggested by the simulations.

 In Figures \ref{fig:T_fb} and \ref{fig:dV_fb}, 
PDFs of $T_{\rm peak}$ and $\sigma_V$
for the feedback and non-feedback area are presented.
 While difference in the $\sigma_V$ PDFs is not clear, 
the $T_{\rm peak}$ PDF for the feedback area is shallower than that for the non-feedback area.
 This indicates that 
an increase of $\sigma_V$ due to feedback is not evident at the current 
\edit1{definition ($r=2''$),}
and that gas in the feedback area tends to be warmer 
and/or a filling factor is larger.
 The former further suggests that the large $\sigma_V$ in the bar 
is not due to stellar feedback.
 To investigate the cause of difference in $T_{\rm peak}$ PDF profiles, 
gas temperature and density need to be measured.
 This requires multi-transition data sets at the same resolution
that will be presented in a future paper.

\section{Molecular gas in M83 bar}\label{sec:discussion}
 From the PDF analysis described above, we find the $I_{\rm CO}$ tail in the bar of M83.
 While such a tail has been regarded as a sign of active star formation in MW studies, 
SFE in the bar of M83 is lower than in the arm.
 We also find a similar tail for $\sigma_V$ in the bar and 
deduce that this large $\sigma_V$ is a major driver for the $I_{\rm CO}$ tail.
 Meanwhile, at least at the current spatial scale, 
stellar feedback does not affect $I_{\rm CO}$ and $\sigma_V$ PDFs.
 This suggests that the large $\sigma_V$ in the bar is not due to feedback.
 It is rather likely that the large $\sigma_V$ suppresses star formation in the bar.

 There are several other studies suggesting an explanation for low SFE in bars 
from a dynamical point of view.
 Based on a large velocity gradient analysis of molecular clouds in Maffei 2,
\citet{Sor12} suggested that clouds in the bar region are likely unbound 
and that the $X_{\rm CO}$ factor can be a factor of 2 smaller than 
that in the arm region.
\edit1{
 If this is also the case for M83, 
the $\Sigma$ PDF for the bar would not have a clear excess 
compared to that for the arm, as the current excess in $I_{\rm CO}$ 
is $\sim 0.5$ dex, i.e., a factor of $\sim 3$.
}
 \citet{Meidt13} presented that a radial area with large streaming motions 
is associated with low SFEs in a spiral galaxy M51.
 They interpreted this result that shearing motions can stabilize molecular clouds.
 These studies support the idea that individual clouds are under a condition 
which is unlikely to form stars.
\edit1{
 From a 2D hydrodynamical simulation for M83, 
\citet{Nimo13} found that clouds in the bar are more virialized 
and SFE should be $\sim 60$\% less in the bar compared to the arm.
}
 On the other hand, 
\edit1{based on a more recent 3D simulation,}
\citet{FujiY14b} proposed a scenario that suppressed star formation can be explained by
high speed cloud-cloud collisions. 

 Since these studies focus on molecular clouds, 
we here briefly summarize cloud properties derived in paper I.
 Note again that region definitions are not identical.
\edit1{
 Compared to clouds in the arm, those in the bar are typically 
larger (radius of $76^{+55}_{-16}$ pc for the bar and $62^{+20}_{-14}$ pc for the arm), 
with higher $T_{\rm peak}$ ($5.2^{+2.4}_{-0.9}$ K for the bar and $3.8^{+1.0}_{-1.1}$ K for the arm), 
with larger $\sigma_V$ ($18.7^{+17.1}_{-4.5}$ km/s for the bar and $11.9^{+3.5}_{-2.9}$ km/s for the arm), 
and with larger virial parameter ($1.6^{+2.1}_{-0.5}$ for the bar and $1.0^{+0.5}_{-0.4}$ for the arm). 
}
 This result is consistent with the idea that cloud property itself is against star formation.
 However, there is a possibility that large clouds \edit1{with large velocity dispersion} 
are a result of cloud-cloud collisions.
 Furthermore, 
\edit1{ 
the differences of the above parameters between the bar and arm are rather small 
and comparable to their dispersions.
}
 For further discussion, again, gas density needs to be derived more accurately
without assuming a constant conversion factor.
 

\section{Summary}

 Based on ALMA and NRO45m observations 
toward M83 in CO(1--0) at 40 pc scale, 
we create PDFs of $I_{\rm CO}$, $T_{\rm peak}$, and $\sigma_V$.
 This work is complementary to paper I, which identifies molecular clouds 
based on the same data set and investigates their properties.

 We first investigate PDF profiles according to the galactic structure, 
i.e.\ center, bar, and arm.
 The $I_{\rm CO}$ PDF for the bar shows an excess or tail at the bright end 
($I_{\rm CO} \ga 100$ K km/s)
compared to that for the arm.
 From PDF studies of MW molecular clouds, such a bright-end tail is 
often interpreted as a sign of self-gravity and/or star formation.
 However, SFE in M83 bar is lower than in arms.
 The $\sigma_V$ PDF for the bar also shows an excess at $\sigma_V \ga 10$ km/s, 
suggesting that this large velocity dispersion in the bar 
is a major cause of the $I_{\rm CO}$ tail 
and low SFE.
 This result indicates that at least at 40 pc scale 
the $I_{\rm CO}$ tail is not always a sign of active star formation.

 We also investigate if stellar feedback affects PDF profiles.
 A feedback area and non-feedback area are defined 
according to catalogs of HII regions, young star clusters, and SNRs.
\edit1{
 The radius of the feedback area is set to be $2''$ ($\sim 40$ pc).
}
 While the $T_{\rm peak}$ PDF is shallower for the feedback area, 
$\sigma_V$ and $I_{\rm CO}$ PDFs do not differ significantly between the two areas.
 We thus conclude that at least at 40 pc scale the feedback effect
does not explain the different PDFs between 
bar and arm regions.

 These results suggest that 
the large-scale disk structures affect PDFs of $\sigma_V$ 
and thus of $I_{\rm CO}$ 
while the small-scale feedback affects the $T_{\rm peak}$ PDF.


\acknowledgments
\edit1{
 We appreciate careful reading and suggestions by a referee 
that have improved our manuscript.
}
 This paper makes use of the following ALMA data: ADS/JAO.ALMA\#2011.0.00772.S.
 ALMA is a partnership of ESO (representing its member states),
NSF (USA) and NINS (Japan), together with NRC (Canada),
NSC and ASIAA (Taiwan), and KASI (Republic of Korea),
in cooperation with the Republic of Chile.
 The Joint ALMA Observatory is operated by
ESO, AUI/NRAO and NAOJ.
 The Nobeyama 45-m radio telescope is operated by Nobeyama Radio Observatory, 
a branch of National Astronomical Observatory of Japan. 

%

\vspace{5mm}
\facilities{ALMA, Nobeyama 45m Telescope}





\appendix

\edit1{
 As mentioned in \S \ref{sec:pdf}, 
$\sigma_V$ (moment 2 of the CO cube) is not corrected for 
the galactic rotation and beam smearing.
 In order to estimate their effect on $\sigma_V$ PDF profiles, 
we have generated a model cube assuming that the disk is super-thin and uniform.
 The model rotation curve in Figure 21 of H14
and the disk orientation parameters in their Table 1 have been adopted.
 The model cube has been smoothed with the synthesized beam mentioned in \S \ref{sec:combine}, 
and then converted to a moment 2 map.
 We have found that the model velocity dispersion is at most 4.3 km/s 
and generally decreases with radius.
 As a result, differences between observed $\sigma_V$ and model-subtracted $\sigma_V$ 
are visible only when $\sigma_V$ is smaller than the channel width (2.6 km/s).
 Since we mostly focus on high-end ($\sigma_V \ga 10$ km/s) profiles of $\sigma_V$ PDFs 
for the bar and arm regions, the effect of the galactic rotation and beam smearing
does not change our conclusions.
 Even for the center region, 
the effect only appears on the low-end profile.
}

\bibliography{papers}



\end{document}